\documentclass[aps,showpacs,twoside,graphixs,epsfig,floatfix]{revtex4}
\usepackage{graphicx}
\usepackage{epsfig}
\def\bra{\langle}
\def\ket{\rangle}
\def\<{\langle}
\def\>{\rangle}
\begin{document}
\title{Negative-Parity States and $\beta$-decays in odd Ho and Dy Nuclei with A=151,153   }
\author{ Falih H. Al-Khudair$^{1,2}$, Gui Lu Long$^{1,2}$
  and  Yang Sun$^3$ }
\address{ $^1$Department of Physics,  Tsinghua University, Beijing 100084, China\\
$^2$ Center of Nuclear Theory, Lanzhou Heavy Ion National
Laboratory, Lanzhou, 730000, China\\
 $^3$ Department of Physics, Shanghai JiaoTong University,
Shanghai 200240, China}
  \date{2007-12-18}
\begin{abstract}
We have investigated the negative-parity states and
 electromagnetic transitions  in $^{151,153}$Ho and $^{151,153}$Dy within the  framework of the
 interacting boson fermion model 2 (IBFM-2). Spin assignments for some states with uncertain
  spin are made based on this calculation.
 Calculated excitation energies, electromagnetic transitions and
 branching ratios are compared with available experimental data and a good agreement is obtained.
 The model wave functions have been used to study $\beta$-decays
 from Ho to Dy isotones, and the calculated $\log ft$
 values are close to the experimental data.

\end{abstract}

\pacs{21.60.Fw, 23.40.Hc, 27.70.+q  }

\maketitle

\section{Introduction}
\label{s1} The interacting boson model(IBM) has been remarkably
successful in describing the collective phenomena observed in  even
-even medium to heavy mass nuclei \cite{Tas1,Sev,Hac}. In the
simplest version of this model, the IBM-1, the nuclear properties
are described by a system of a fixed number of boson. In this
version no distinction is made between proton boson and neutron
boson, therefore all states in IBM-1 are $F$-spin symmetric
\cite{Ari1,Ari2,Ari3}. The building blocks  are  $ (s^{\dagger},
\tilde{s}) $ for s-boson and $ (d^{\dagger},\tilde{d~}) $ for
d-boson. The second version,  the IBM-2, does distinguish  between
proton boson and neutron boson. The states in IBM-2 include all the
$F$-spin symmetric  states as well as mixed symmetry states
belonging to the U(6) representation $[N-1,1]$. An important
property of this new version is that the proton - neutron symmetry
character of each states is specified in terms of a new quantum
number called $F$-$spin$ \cite{Iac1,Ham,Garr,Gade,Elha}. For lighter
nuclei, the IBM has been extended to the interacting boson model
with isospin (IBM-3) \cite{Ell3}.  Within  the IBM-3, the
neutron-proton pair must be included in addition to the two other
types of bosons in the IBM-2, and they form an isospin triplet
 \cite{Ell1,Eva1,Garc,FalG,faPRC75}.
 In the interacting boson fermion model
(IBFM) \cite{IachPRL}, odd-A nuclei are described by coupling the
degrees of freedom of odd particle to  a core which is described  in
the IBM. Calculations of positive and negative-parity states and the
electromagnetic transitions of odd mass nuclei have performed within
the framework of IBFM, for instance in Refs.
 \cite{CunNPA385,AriNPA445,LheZPA337,YoshNPA619,BraPRC58,YazPRC75}.
One of most important property in nuclear structure study is the
$\beta$ decay rates.
 The $\beta$ transition for odd-nuclei has received  intensive interests in the last few years
  \cite{Sarr,Hari,Towner,Borzov,Chatu,Lher,Musto,Fisch}.
   Theoretical contribution to the   study of nuclear beta decay
   have been made over the years using the IBFM
     \cite{YoshPRC66,ZufPRC68,BraPRC70},
   and good agreement has been found with  available experimental
   data.

The purpose of the present work is to investigate the
negative-parity states and  electromagnetic transitions in the
$^{151,153}$Ho and $^{151,153}$Dy isotopes by using the IBFM-2
model.  More importantly,  $\beta$ decay between the levels are
studied by using the wave functions obtained from the structure
calculation of this model. In particular, the influence of different
values of hamiltonian parameters on the energies and decay
probabilities is investigated.

In order to calculate an odd-nucleus in the IBFM-2 model, we need to
choose an  even-even core. Here, the even-even $^{150,152}$Dy nuclei
have been chosen as the respective core for $^{151,153}$Dy isotopes. They
 have $66$ protons in
the $50-82$ shell  and $84$ and $86$ neutrons in shell $82-126$,
respectively. Both nucleons lie in the first half shell, therefore
they should be considered as particle bosons. For $^{151,153}$Ho
isotopes, we considered them as    resulting from coupling a proton
hole to the even-even Er nuclei.

In section \ref{s2},
 we briefly review the interacting boson fermion  model.
 In section \ref{s3}, we present our calculation
results  for the energy levels of  the core nuclei and compared with
available
 data. The negative-parity states of $^{151,153}$Ho and
$^{151,153}$Dy nuclei are presented in sections \ref{s4} and
\ref{s5}, respectively. A discussion of  electromagnetic
 transitions follows in section \ref{s6}. In section \ref{s7} the
 $\beta$ decay from levels of odd-proton Ho-isotopes to levels in odd-neutron
 Dy-isotopes levels are studied. Finally, in section \ref{s8} we summarize our results.

\section{The IBFM-2 model}
\label{s2}

The low lying levels in odd nuclei are described as combined system
of a group of bosons with one fermion. In general, the Hamiltonian
for this coupled system can be written as \cite{ZufPRC68},
\begin{equation}\label{e1}
    H= H_{B}+ H_{F}+ V_{BF}.
\end{equation}
Here $H_{B}$ is the usual IBM-2 Hamiltonian which describes the
system of $(s_{\nu},s_{\pi})$ and $(d_{\nu},d_{\pi})$-bosons
\begin{equation}\label{e2}
  H=\varepsilon_{d}(\hat{n}_{d\pi}+\hat{n}_{d\nu})+\kappa_{\pi\nu} \hat{Q}_{\pi}\cdot
  \hat{Q}_{\nu}
   +\sum_{\rho=\pi,\nu}\hat{V}_{\rho\rho}+ \hat{M}_{\pi\nu},
\end{equation}
where  $ \varepsilon_{d} $ is the d-boson excitation  energy,  and $
n_{d\pi}$ and $ n_{d\nu}$  are the neutron and proton $d$-boson
number operator respectively. $ \kappa_{\pi\nu} \hat{Q}_{\pi}\cdot
\hat{Q}_{\nu}$  is the quadruple interaction between proton and
neutron boson, and $ \hat{Q}_{\rho}$, the quadruple operator, is
given by
\begin{equation}\label{e3}
\hat{Q}_{\rho}=(s^{\dagger}_{\rho}\tilde{d~_{\rho}}+
s^{\dagger}d^{\dagger}_{\rho})^{2}+\chi_{\rho}(d^{\dagger}\tilde{d})^{2}.
\end{equation}
The Majorana interaction is
\begin{equation}\label{e4}
 \hat{M}_{\pi\nu}=\xi_{2}[(d^{\dagger}_{\nu}s^{\dagger}_{\pi}-d^{\dagger}_{\pi}s^{\dagger}_{\nu}).
 (\tilde{d~}_{\nu}s_{\pi}-\tilde{d~}_{\pi}s_{\nu})
]^{(2)}+\frac{1}{2}\sum_{k=1,3}\xi_{k}[d^{\dagger}_{\nu}d^{\dagger}_{\pi}]^{(k)}.[\tilde{d}_{\pi}
\tilde{d}_{\nu}]^{(k)},
\end{equation}
and it only affects the positions of the mixed symmetry states. The
$ V_{\rho\rho} $ term represents the interaction between like
bosons,
\begin{equation}\label{e5}
  \hat{V}_{\rho\rho}=1/2\sum_{L=0,2,4}[2L+1]C_{\rho}^{(L)}[d_{\rho}^{\dagger}d_{\rho}^{\dagger}]^{(L)}.
  [\tilde{d~}_{\rho}\tilde{d~}_{\rho}]^{(L)},
\end{equation}
   where $\rho=\pi,\nu$.

The term $H_{F}$ is the Hamiltonian of odd fermion,
\begin{equation}\label{e6}
  H_{F}=\sum_{i}\epsilon_{i}n_{i},
\end{equation}
where $\epsilon_{i}$ is the quasi-particle  energy of the $i$th
orbital, and  $ n_{i}$ is   fermion number operator. The
quasi-particle energies and occupation probabilities  are usually
calculated using the \textbf{BCS}  \cite{BardenPR108} approximation
in terms of the Fermi energy $\lambda$, the paring gap $\Delta$ and
the single-particle energies $E_{i}$,
\begin{equation}\label{e7}
  \epsilon_{i}=\sqrt{(E_{i}-\lambda)^{2}+\Delta^{2}}.
\end{equation}
The occupation probabilities are then given by
\begin{equation}\label{e8}
  \upsilon_{i}=[\frac{1}{2}(1-\frac{E_{i}-\lambda)}{\epsilon_{i}})]^{1/2},\hspace{.8in}
  u_{i}=(1-\upsilon_{i}^{2})^{1/2}.
\end{equation}
 The bosons-fermion
interaction $V_{BF}$ is, in general, rather complicated but it has
been shown \cite{ZufPRC68} to  be dominated by  three terms,
\begin{equation}\label{e9}
  V_{BF}=\sum_{i}A_{i} n_{i}n_{d_{\acute{\rho}}}+\sum_{i,j}\Gamma_{ij}([a^{\dagger}_{i}\tilde{a}_{j}]^{(2)}\cdot
  Q^{B}_{\acute{\rho}})+
  \sum_{i,j}\Lambda^{j}_{ki}\{:[[d^{\dagger}_{\rho}\tilde{a}_{j}]^{(k)}a^{\dagger}_{i}s_{\rho}]^{(2)}: \cdot
  [s^{\dagger}_{\acute{\rho}}\tilde{d}_{\acute{\rho}}]^{(2)}+
  H.c.\},
\end{equation}
where $\tilde{d_{\mu}}=(-1)^{\mu}d_{-\mu}$,
$\tilde{a}_{j\mu}=(-1)^{j-\mu}a_{j-\mu}$ and $Q^{B}_{\rho}$ is the
boson quadrupole operator which is defined in  eq. (\ref{e3}). The
symbols $\rho$ and $\acute{\rho}$ denote $\pi (\nu)$ and $\nu (\pi)$
if the odd fermion is a proton (neutron). The first
 term in Eq.(\ref{e9}) is a monopole-monopole interaction, which is
proportional to the number of d-bosons. Therefore it only gives
rise to a renormalization of the boson energy $\varepsilon =
\varepsilon_{d}- \varepsilon_{s}$ and it does not affect the main
structure of energy spectrum. The second term
 is a quadrupole-quadrupole interaction, and the last term is the
 exchange interaction. The origin of the exchange force is closely related to the presence of the Pauli principle.
   Both terms are  dominant terms and appear to arise from the strong
 neutron-proton quadrupole force. The orbital dependence of the coupling coefficients has been
 microscopically estimated  \cite{Schoinbook}
\begin{eqnarray}
  \Gamma_{i,j}&=&(u_{i}u_{j}-\upsilon_{i}
  \upsilon_{j})Q_{i,j}\Gamma,\label{e10}\\
  \Lambda^{j}_{k,i}&=&-\beta_{k,i}
  \beta_{j,k}(\frac{10}{N_{\rho}(2j_{k}+1)})^{1/2}
  \Lambda,\label{e11}\\
  \beta_{i,j}&=&(u_{i}u_{j}+\upsilon_{i}
  \upsilon_{j})Q_{i,j},\label{e12}\\
 Q_{i,j}&=&\langle l_{i},\frac{1}{2},j_{i}\|Y^{(2)}\|
 l_{j},\frac{1}{2},j_{j}\rangle.\label{e13}
\end{eqnarray}

\section{ even-even nuclei  Structure}
 \label{s3}
\begin{figure}[htp]
\begin{center}
\includegraphics[width=6in,height=5in]{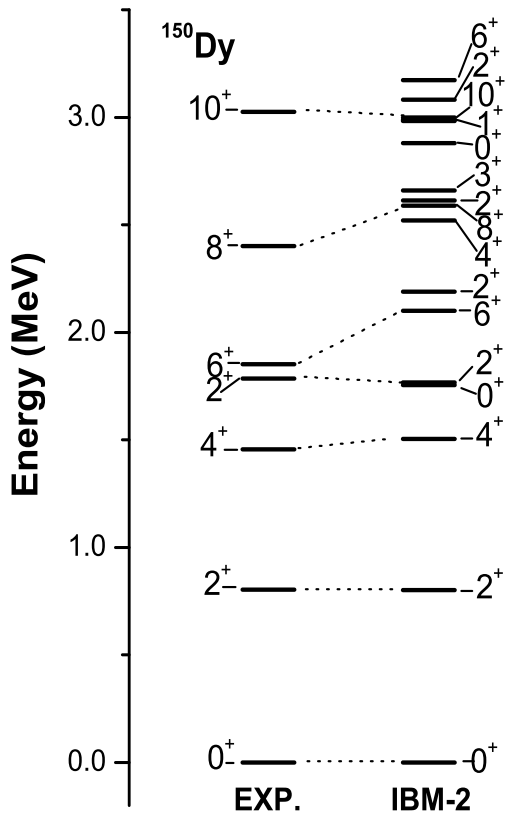}
\caption { The calculated and  observed  energy  spectra for the
$^{150}$Dy  isotope.}\label{f1}
\end{center}
\end{figure}

\begin{figure}[htp]
\begin{center}
\includegraphics[width=6in,height=5in]{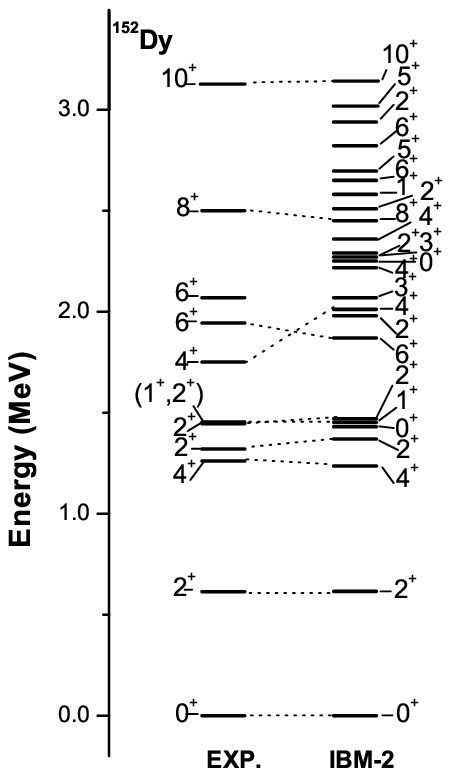}
\caption { The calculated and  observed  energy  spectra for the
$^{152}$Dy  isotope.}\label{f2}
\end{center}
\end{figure}

\begin{figure}[htp]
\begin{center}
\includegraphics[width=6in,height=5in]{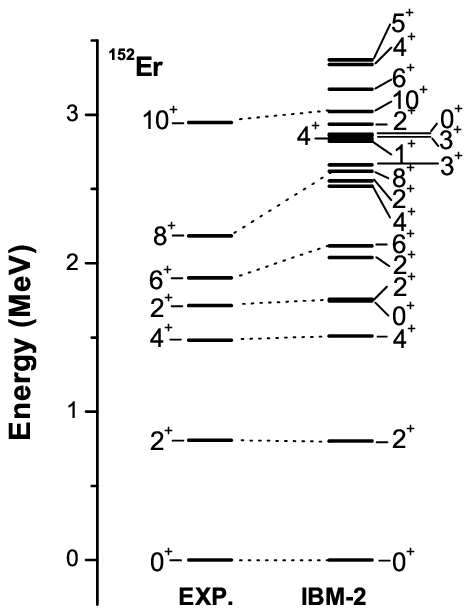}
\caption { The calculated and  observed  energy  spectra for the
$^{152}$Er  isotope.}\label{f3}
\end{center}
\end{figure}

\begin{figure}[htp]
\begin{center}
\includegraphics[width=6in,height=5in]{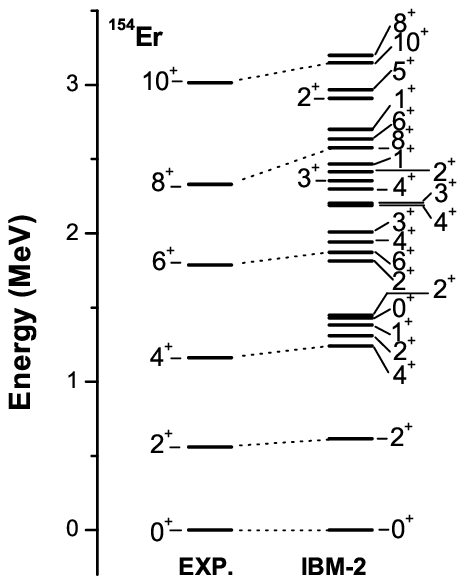}
\caption { The calculated and  observed  energy  spectra for the
$^{154}$Er  isotope.}\label{f4}
\end{center}
\end{figure}

In the calculation the proton and neutron shells are assumed to
closed at Z=50  and N=82 magic shells. For $^{150,152}$Dy nuclei,
there are eight proton-bosons  and one and two neutron-bosons,
respectively. On  the other hand, there are seven hole-like proton
bosons
 and one and two particle-like neutron bosons  for  $^{152,154}$Er nuclei, respectively.
    The calculated excitation energies are obtained by
diagonalizing the Hamiltonian in Eq. (\ref{e2}), usually using the
NPBOS code \cite{Ost1}. The parameters $\varepsilon_{d} $ and
$\kappa_{\pi\nu} $ have been determined so as to reproduce as
closely as possible the energy of the low lying  positive parity
states. The energy of the ground state band levels were
   optimized by varying the proton anharmonicity parameters
   $C_{\pi}^{L}(L=0,2,4 )$. The  parameters  $ \chi_{\pi} $ and  $ \chi_{\nu} $  have
 been kept constant in the two isotopes, and are taken as the
 same as those for the SU(3) limit: $ \chi_{\pi} $ =  $ \chi_{\nu}
 $ = $-\sqrt{7}/2$. In order to identify mixed  symmetry states,  we fitted the energy
 of all the $ J=2^{+}$ states  below 3 MeV by smoothly changing  $\xi_{2} $.
 The best fit values for the Hamiltonian parameters are given in
 Table \ref{t1} and the calculated energy levels are compared with
 available experimental data as shown in Figs.  \ref{f1}-\ref{f4}. Good agreement   between the calculated and observed
spectrum is obtained for the ground state band. From Fig. \ref{f1}
one can see that the sequence of levels is well reproduced, though
the calculated excitation energies of $ 6_{1} ^{+}$ and $8_{1} ^{+}$
are somewhat higher than the experimental ones. The splitting of $
2_{2}^{+} $ and $4_{1}^{+} $  in the "two phonon states"  is well
reproduced, and it justifies the value of $C_{\pi}^{L}$ that are
used. The energy of $ 0_{2}^{+}$ " of the two phonon states" equals
to 1.757 MeV in the IBM-2, and this remains to be seen in future
experiment.

In $^{152}$Dy nucleus, the level at $ 1.452$ MeV have possible
 $J=1^{+},2^{+}$  assignments in experiment. It  is  close to a level
 at $ 1.452 $ MeV with $J=1^{+}$ in our IBM-2 result.
The $4^{+}_{2}$ state is higher than the data. There is no suitable
solution in the present scheme for this problem. One possible
explanation is the effect of the g-boson. The second $J=2^{+}$ state
at energy $1.379$ MeV   is  close to the experimental level at 1.313
MeV. The calculated   $J=2_{3} ^{+}$ state at  $1.470$ MeV   is in
good agreement with the experimental level at $1.448$ MeV,   and
this state has  mainly the $F=F_{max} $ component. For
$^{152,154}$Er nuclei, the ground state band fits well with data,
although we  used the same parameters   that adjusted   for
$^{150,152}$Dy nuclei as shown in Figs. \ref{f3} and \ref{f4}. In
$^{152}$Er, the calculated  $J=2_{2} ^{+}$ state at 1.758 MeV is
close to  the experimental one at 1.715 MeV, and this state has  a
very pure $F_{max}$ character. The calculation result shows that the
wave function for $J=2_{2} ^{+}$ is composed of 100\% of  the $
s^{N-2}d^{2}$ configuration.

\section{Odd-proton $^{151,153}$Ho Structure }
 \label{s4}

\begin{figure}
\begin{center}
\includegraphics[width=7in,height=6in]{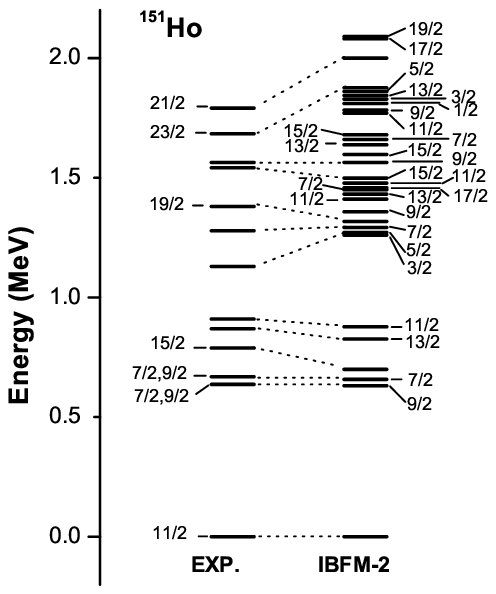}
\caption { The calculated and  observed  energy  spectra for the
$^{151}$Ho isotope.}\label{f5}
\end{center}
\end{figure}

\begin{figure}
\begin{center}
\includegraphics[width=7in,height=6in]{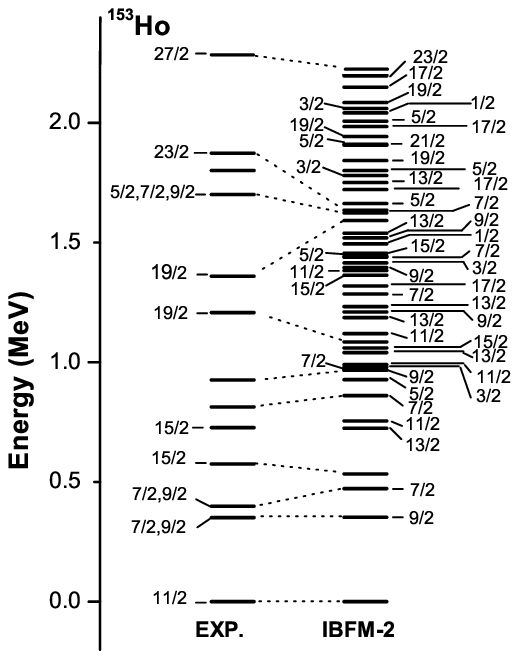}
\caption { The calculated and  observed  energy  spectra for the
$^{153}$Ho isotope.}\label{f6}
\end{center}
\end{figure}

 In order to study especially the influence of fermion degrees of freedom,
  we have investigated odd-neutron Dy nuclei and odd-proton  Ho
  nuclei with the same boson parameters  for each odd-A number. Thus, the differences in the nuclear
  structures of these two nuclei can arise only from  the boson-fermion  interaction
  and the  odd-particle Hamiltonian. The quasi-particle  energies $\epsilon_{i}$ and occupation
probabilities $ \upsilon ^{2}$ are determined  from a simple
\textbf{BCS} calculation using Reeehal and  Sorensen
 \cite{ReehPRC2} single-particle energies. The single-particle
energies  taken are given in Table \ref{t2}.  The \textbf{BCS}
equations are resolved with the single-particle orbitals $ g_{9/2}$,
$ g_{7/2}$, $d_{5/2}$, $h_{11/2}$, $d_{3/2}$ and $s_{1/2}$, and with
$\Delta $= 12$/\sqrt{A}$.  In the description of negative-parity
states in Ho-isotopes,  the odd proton is taken to be in the $
h_{11/2}$ orbital. We searched  the optimal values  of the
interaction parameters that describe well the experimental  levels
and electromagnetic transitions. The found values of the parameters
are shown in Table \ref{t2}. In Ho-isotopes which have $67$ protons,
the occupation probabilities of $\pi h_{11/2}$ is $
\upsilon^{2}\simeq 0.32$. As a result, the parameter of exchange
force in $V_{BF}$ plays a crucial role in fitting the experimental
data. The strength of the exchange force in the $V_{BF}$ has to be
increased  in order to lower the energy of the first $
\frac{9}{2}^{-}$ state in $^{153}$Ho. The $\Gamma $ was kept
constant for both isotopes.  The calculated and observed  \cite{ENSDF} energy spectra of $^{151,153}$Ho are shown in
 Figs.
 \ref{f5} and \ref{f6} respectively. The calculation gives a number of predictions, and they
  are presented in the figures which are helpful to future experiments.
 From the figures, we see  that the calculated energy levels agree
  with the experimental data quite well in general. Reproduction of the  trend in
 the experimental data is clear, especially those of the first and second appearance of   negative-parity states.

 Experimentally, the first and second  negative parity  excited states have possible
 $J=(\frac{9}{2}^{-},\frac{7}{2}^{-})$  assignments, and they  are  close to
 states with  $J=\frac{9}{2}^{-}$ and $\frac{7}{2}^{-}$, respectively in the IBFM-2 calculation for
  both isotopes.
 In $^{151}$Ho the states  at $ 0.869 $ and $0.910$ MeV in the  experimental
  data are  close to  the  states  with  $ J= (\frac{13}{2}^{-})_{1} $ and
 $ J= (\frac{11}{2}^{-})_{2} $ in the
IBFM-2 results at $0.826$ and $0.878$ MeV, respectively. The
observed order   inversion in Ho-isotopes, namely
$\frac{21}{2}^{-}$-$\frac{23}{2}^{-}$ has also been reproduced.
However, the calculated energy of the $ (\frac{21}{2}^{-})_{1} $ and
$ (\frac{23}{2})_{1}$ state in $^{151}$Ho are larger than the
experimental value. This is probably due to the restriction of the
limited  single-particle space in the $h_{11/2}$ orbital
\cite{CunNPA385}.

In $^{153}$Ho the calculated   $(\frac{15}{2}^{-})_{1}$ state
  at 0.534 MeV is close to the  experimental  level at 0.576 MeV, while the IBFM-2 calculation gives the
  $(\frac{15}{2}^{-})_{2}$ at 1.046 MeV and it is   far from the
  experimental  one   at 0.727 MeV. On other hand, the energy of the
$(\frac{13}{2}^{-})_{1}$ state in the  model calculations equals to
0.725 MeV. We have considered two possible theoretical assignments
for the experimental state at 0.926 MeV: $\frac{5}{2}^{-}$ and
$\frac{9}{2}^{-}$. The calculations indicate that the
$(\frac{9}{2}^{-})_{1}$ is more probable.  The characteristics of
the experimental level at 1.700 MeV with possible
$(\frac{5}{2}^{-},\frac{7}{2}^{-}, \frac{9}{2}^{-})$ assignment is
in good agreement  with our calculated  $\frac{7}{2}^{-}$ state at
1.624 MeV.

The higher spin states, such as $\frac{25}{2}^{-}$ and
$\frac{27}{2}^{-}$  of $^{153}$Ho, have also been reproduced. The
calculated energies for the $\frac{25}{2}^{-}$ and
$\frac{27}{2}^{-}$ states in $^{153}$Ho are 2.490 and 2.201 MeV,
respectively, which are close to experimental levels at 2.358 and
2.297 MeV respectively.

\section{Odd-neutron $^{151,153}$Dy structure }
 \label{s5}

 For Dy-isotopes, the \textbf{BCS}  equations  are solved with the
single-particle orbitals  $ f_{7/2}$, $h_{9/2}$, $p_{3/2}$,
$f_{5/2}$, $ i_{13/2}$, $ h_{11/2}$ and $p_{1/2}$ with $\Delta $=
12$/\sqrt{A}$. The values of single-particle energies are extracted
from Ref. \cite{Bohbook}, and they are very similar  to the ones
used in Ref. \cite{BijNPA379}, except  for the $ i_{13/2}$ orbital.
In our calculations the $ i_{13/2}$  is still present between the
$h_{9/2}$ and  $p_{3/2}$ orbitals but close to $p_{3/2}$ orbital in
  both isotopes. The adopted single-particle energies are listed
in Table \ref{t3}.
 In $^{151,153}$Dy-isotopes, the $\nu p_{1/2}$ level is almost
 completely empty $(\upsilon ^{2}\approx 0)$ while the  $\nu h_{11/2}$ level is almost
 completely  occupied   $(\upsilon ^{2}\approx 1)$,  thus  the
 two orbitals were omitted from the calculation. According to
 this result,  we include only the first four orbitals in calculating the negative-parity
states.

\begin{figure}
\begin{center}
\includegraphics[width=7in,height=6in]{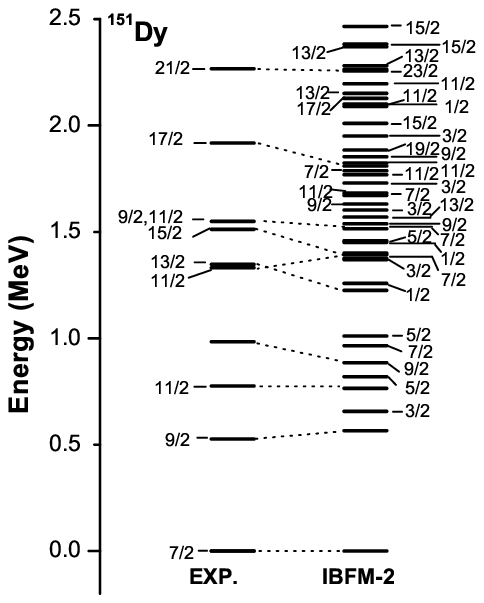}
\caption { The calculated and  observed  energy  spectra for the
$^{151}$Dy isotope.}\label{f7}
\end{center}
\end{figure}

\begin{figure}
\begin{center}
\includegraphics[width=7in,height=6in]{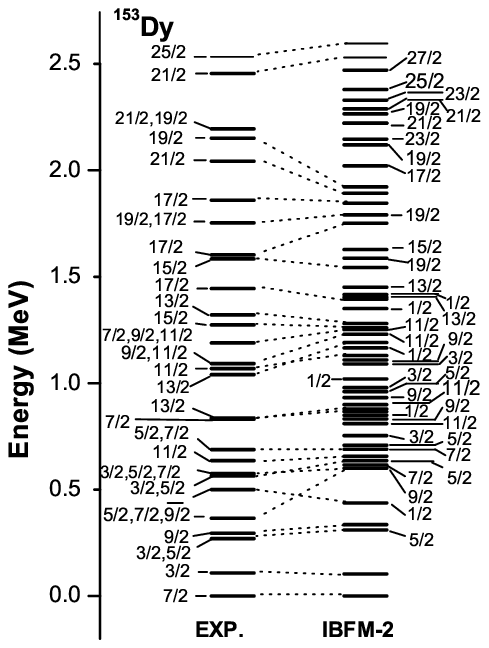}
\caption { The calculated and  observed  energy  spectra for the
$^{153}$Dy isotope.}\label{f8}
\end{center}
\end{figure}

A comparison of the results of our IBFM-2 calculations with the
experimental results  \cite{ENSDF} on the low-lying negative-parity
states of $^{151,153}$Dy isotopes is shown in Figs. \ref{f7} and
\ref{f8}, respectively. All levels below 2.5 MeV known from
experiment are included. From Fig. \ref{f7}, one can see that the $(
\frac{9}{2}^{-})_{1}$ and $( \frac{11}{2}^{-})_{1}$ states agree
very well between calculation and experiment. For $^{151}$Dy the
experimental data up to $ 2.5$ MeV is scarce. The low-lying state
with $0.984$ MeV  has not been assigned any other quantum number. It
has a transition $E_{\gamma}= 0.209$ and $\log ft$ = $5.8$ from
$^{151}$Ho ground state $\frac{11}{2}^{-}$. From our calculation, it
 is close to the calculated state at $0.886$ MeV with $J^{-} =(
\frac{9}{2}^{-})_{2}$. This assignment is further enforced by our
$\log ft$ studies to be presented shortly.

Another state with uncertain spin assignment
$(\frac{9}{2},\frac{11}{2}^{-})$ is  at an excitation energy of
$1.549$ MeV with $\gamma$-transition  to $(\frac{7}{2}^{-})_{1}$
state,  and it has a $\log ft$= $5.1$ from $^{151}$Ho ground state
$\frac{11}{2}^{-}$. It is reproduced very well in our IBFM-2
calculation with an excitation energy of  $1.523$ MeV with $J^{-} =
(\frac{9}{2}^{-})_{3}$.  There is not $J^{-} = \frac{11}{2}^{-}$
state in energy range between 1.40 MeV  and 1.68 MeV in our
calculation. Our calculation predicts  $(\frac{3}{2}^{-})_{1}$ and
$(\frac{3}{2}^{-})_{2}$ at 0.657 MeV and 1.369 MeV   in $^{151}$Dy,
and they have  not been observed experimentally, while in $^{153}$Dy
the $(\frac{3}{2}^{-})_{1}$ state is  at 0.104 MeV in our IBFM-2
calculation, and this very well reproduced the experimental level at
  0.108 MeV.

In Fig. \ref{f8} we present a more detailed comparison between
experimental and calculated  energy states in $^{153}$Dy. Because
many states  have no clear assignment, so special attention is given
to these levels with the hope to give assignment to them. Indeed, it
is found that many calculated states are quite close to the
experimental ones, and it will be highly desirable to substantiate
this model prediction in future experiment. A detailed presentation
is given in Table \ref{t4} where we list the calculated energy
levels, available experimental assignments ( certain and uncertain
spin assignment).

 We also analyzed the single particle occupation probability for
 interested states in
Dy-isotopes, and they   are summarized in Table \ref{t5}. It is
apparent that for the first few states, they are mainly single
quasi-particle excited states where one of the occupation
probability is dominant. As we go to higher excited states, we see a
spreading of the occupation into more single particle states.  The
most important single-particle orbitals are those closest to the
Fermi level, and they are  $f_{7/2}$ and $h_{9/2}$. Clearly, for
$^{153}$Dy, the $\frac{21}{2}^{-}$ state is lower than that of the
$\frac{23}{2}^{-}$ in energy, and there is no order inversion here.

\section{Electromagnetic Transitions}
   \label{s6}

   In the IBFM-2 model the electromagnetic transition operator
   is described by the following operator
\begin{equation}\label{e14}
T^{(\lambda)}=T^{(\lambda)}_{B}+ T^{(\lambda)}_{F},
\end{equation}
which contains a boson part and fermion part. The $E2$ transition
operator is expressed \cite{ZufPRC68}
 \begin{equation}\label{15}
        T^{E2} =  e^{B}_{\pi} Q^{B}_{\pi}+e^{B}_{\nu}
        Q^{B}_{\nu}+
        \sum_{i,j}e^{(2)}_{i,j}[a^{\dagger}_{i}\tilde{a}_{j}]^{(2)},
        \end{equation}
where  the quadrupole operators $Q_{\pi}$ and $Q_{\nu}$ are defined
in  Eq.(\ref{e3}), $ e_{\pi} $ and $ e_{\nu}$ the proton and neutron
boson effective charges and
\begin{equation}\label{16}
       e^{(2)}_{i,j}=-\frac{1}{\sqrt{5}}(u_{i}u_{j}-\upsilon_{i}\upsilon_{j})\langle l_{i},\frac{1}{2},j_{i}\|r^{2}Y^{(2)}\|
 l_{j},\frac{1}{2},j_{j}\rangle.
        \end{equation}
 The $M1$ transition operator in  IBFM-2 is
\begin{equation}\label{17}
        T^{M1} =  \sqrt{\frac{3}{4\pi}}(g^{B}_{\pi} L^{B}_{\pi}+g^{B}_{\nu}
        L^{B}_{\nu}+\sum_{i,j}e^{(1)}_{i,j}[a^{\dagger}_{i}\tilde{a}_{j}]^{(1)}),
        \end{equation}
where  $g_{\pi} $ and $g_{\nu}$ are g factors for proton and neutron
boson , $\tilde{L}$ is the angular momentum operator
\begin{equation}\label{18}
L_{\rho}=\sqrt{10}[d^{+}_{\rho}\tilde{d}_{\rho}]^{(1)},
\end{equation}
and the coefficient
\begin{equation}\label{19}
e^{(1)}_{i,j}=-\frac{1}{\sqrt{3}}(u_{i}u_{j}+\upsilon_{i}\upsilon_{j})\langle
l_{i},\frac{1}{2},j_{i}\|(g_{l}\textbf{l}+g_{s}\textbf{s})\|
 l_{j},\frac{1}{2},j_{j}\rangle,
\end{equation}
where $g_{l}$ and $g_{s}$ are the single particle g-factors of the
odd nucleon. For boson part, the E2 matrix elements are very
sensitive to the difference between neutron boson and proton
effective charge, and they are  kept constant  at
$e_{\pi}=2e_{\nu}=0.1$ $ e.b $  for all isotopes. The values of
$e^{B}_{\rho}$ were determined from the experimental
$B(E2;2^{+}_{1}\rightarrow 0^{+}_{1})$ of even-even $^{152}$Dy
nucleus. For the odd nucleon, the effective charge $1.5$ $e$ and
$0.5$ $e$ are taken for the proton and the neutron, respectively.
The parameter $ \chi$ in the $E2$ transition operator has the same
value as in the Hamiltonian, though they are not necessary
\cite{long1,long2}. The standard boson $g$ factor values $g_{\pi}$ =
1 $\mu_{N} $ and $g_{\nu}$ = 0 $\mu_{N} $ are used for all isotopes.
We have estimated the  single particle  $g_{l}$ and $g_{s}$, and
taken them as $g_{l,\nu}^{F}$ = 0 $ \mu_{N}$ and $g_{l,\pi}^{F}$ = 1
$ \mu_{N}$, while the spin $g$-factors are taken as the free values
quenched by a factor of 0.7 and 0.5 for proton and neutron,
respectively, which is the common practice as in Refs. \cite{
AriNPA445,Alonso,GiZon}. Using this procedure we have calculated the
electromagnetic transitions, and  a very good agreement between
calculated and experimental magnetic moment of the ground states is
obtained in both magnitude and sign as shown in Tables \ref{t10} and
\ref{t11}.

 The resulting branching ratios
for the $ ^{151,153}$Ho and $ ^{151,153}$Dy nuclei are listed in
Tables \ref{t6}-\ref{t9}, respectively, in comparison with the
experimental data  \cite{ENSDF}. In these tables, the strongest
branch is correctly predicted. The other transitions are in
qualitative agreement with experiment. The deviations can be reduced
by changing $g_{s}$, as well as by changing the effective charges.
From Tables \ref{t5},\ref{t8} and \ref{t9}, one can see that  the
strongest transitions are between those states having the same
dominated single-particle orbital. This means that a strong
transition should occur between levels in the same band. Our results
for electromagnetic transition probabilities are summarized in
Tables \ref{t10} and \ref{t11} respectively. In $^{151,153}$Dy  the
$ \frac{5}{2}^{-}_{1}$ states decay predominantly  to the $
\frac{3}{2}^{-}_{1}$ states via a pure M1 transition. In$^{153}$Dy
isotope, the $ \frac{5}{2}^{-}_{1}$ is associated with $\nu f_{7/2}$
and $\nu h_{9/2}$, the exhibit sizable mixing of these two
quasi-particle orbitals. Due to the mixing of the two components in
the wave function in this state,  the $ \frac{5}{2}^{-}_{1}$ and $
\frac{7}{2}^{-}_{1}$ states are connected by  strong  E2 transition
and  weak M1 transition in $^{153}$Dy isotope. In contrast, we see
that the $ \frac{5}{2}^{-}_{1}$ state decays to $
\frac{7}{2}^{-}_{1}$
 states by very  strong  E2  and   M1 transitions in
 $ ^{151}$Dy  isotope.

We have also calculated the quadrupole moments of the ground states
and some low-lying states  in $ ^{151,153}$Ho and $ ^{151,153}$Dy.
The results are given in Tables \ref{t10} and \ref{t11}
respectively. The calculated results are in the same order of
magnitude as the available experimental data.

\section{$\beta$-decay}
\label{s7}

In IBFM-2, a relation among the IBM and the underlying shell model
has been established by including the proton and neutron degree of
freedom \cite{Iac1}. This offers  one the capability to compute the
probabilities of $\beta$-decay. The decay of odd-nuclei proceeds
predominantly through the conversion of the odd particle from
neutron to proton ($ \beta^{-}$-decay) or from proton to neutron ($
\beta^{+}$-decay). There are two types of beta decay, the Fermi
decay and Gamow-Teller decay. In the framework of IBFM  both
transitions can be calculated \cite{Dellagiacoma}. First define the
following operators,
\begin{eqnarray}
A^{\dagger(j)}_{m}&=&\zeta_{j}a^{\dagger}_{jm}+\sum_{\acute{j}}\zeta_{j\acute{j}}s^{\dagger}
[\tilde{d}a^{\dagger}_{\acute{j}}]^{(j)}_{m} \hspace{.2in}(\Delta
n_{j}=1,\Delta N=0),\label{20}\\
B^{\dagger(j)}_{m}&=&\Theta_{j}s^{\dagger}\tilde{a}_{jm}+\sum_{\acute{j}}\Theta_{j\acute{j}}
[d^{\dagger}\tilde{a}_{\acute{j}}]^{(j)}_{m} \hspace{.2in}(\Delta
n_{j}=-1,\Delta N=1),\label{21}\\
\tilde{A}^{(j)}_{m}&=&
(-1)^{j-m}\{A^{\dagger(j)}_{-m}\}^{\dagger}=\zeta^{*}_{j}\tilde{a}_{jm}+\sum_{\acute{j}}\zeta^{*}_{j\acute{j}}s
[d^{\dagger}\tilde{a}_{\acute{j}}]^{(j)}_{m} \hspace{.2in}(\Delta
n_{j}=-1,\Delta N=0),\label{22}\\
\tilde{B}^{(j)}_{m}&=&(-1)^{j-m}\{B^{\dagger(j)}_{-m}\}^{\dagger}=-
 \Theta^{*}_{j}s a^{\dagger}_{jm}-\sum_{\acute{j}}\Theta^{*}_{j\acute{j}}
[\tilde{d}a^{\dagger}_{\acute{j}}]^{(j)}_{m} \hspace{.2in}(\Delta
n_{j}=1,\Delta N=-1).\label{23}
\end{eqnarray}

Then the Fermi and the Gamow-Teller transition operators are
\begin{eqnarray}
Q^{F}&=&\sum_{j}-\sqrt{2j+1}[P^{(j)}_{\pi}P^{(j)}_{\nu}]^{(0)},\label{24}\\
Q^{GT}&=&\sum_{\acute{j}j}\eta_{\acute{j}j}[P^{\acute{(j)}}_{\pi}P^{(j)}_{\nu}]^{(1)},\label{25}
\end{eqnarray}
where
\begin{equation}\label{26}
\eta_{j\acute{j}}=-\frac{1}{\sqrt{3}}\langle
\acute{l},\frac{1}{2},\acute{j}\|\sigma\|
 l,\frac{1}{2},j\rangle
 =-\delta_{\acute{l}l}\sqrt{2(2\acute{j}+1)(2j+1)}W(l\acute{j}\frac{1}{2}1;\frac{1}{2}j).
 \end{equation}
The form of transfer operator  $ P^{j}_{\rho}$ depends  on the
specific nuclei, and in the present case,  $P^{(j)}_{\pi}
=\tilde{A}^{(j)}_{\pi}$ and $P^{(j)}_{\nu} =\tilde{B}^{(j)}_{\nu}$.
The $ft$ value is calculated by
 \begin{equation}\label{27}
ft=\frac{6163}{\bra M_{F} \ket ^{2}+(G_{A}/G_{V})^{2}\bra M_{GT}\ket
^{2}}
\end{equation}
in units of second where $G_{A}/G_{V})^{2}$=1.59, and
\begin{eqnarray}
\bra M_{F}\ket ^{2}&=&\frac{1}{2I_{i}+1}| \bra
I_{f}\|Q^{F}\|I_{i}\rangle|^{2},\label{28}\\
 \bra M_{GT}\ket ^{2}&=&\frac{1}{2I_{i}+1}|\bra
I_{f}\|Q^{GT}\|I_{i}\rangle|^{2}.\label{29}
\end{eqnarray}
 Having obtained the wave functions, we can
calculate the $\beta$-decay  rates. It should be stressed that there
is no adjustable parameters in the beta decay calculation,
consequently the $\log_{10}ft$ is obtained in a parameter free
manner. Examining the wave function of daughter nuclei
$^{151,153}$Dy, the first and second excited $\frac{9}{2}^{-}$
states are dominated by $h_{9/2}$ and $f_{7/2}$ orbitals in both
isotopes, respectively. According  to these components the
$\log_{10}ft$ values of the two states have the approximately the
same values and in good agreement with experimental ones. An
interesting result of the calculation is that it indicates the
observed state at 1.549 MeV in $^{151}$Dy, corresponds to the state
$(\frac{9}{2}^{-})_{3}$ at 1.523 MeV  in the IBFM-2 result, and the
main single-particle component of this state is $h_{9/2}$. Because
 the $(\frac{9}{2}^{-})_{3}$ state in$^{153}$Dy  at
0.862 MeV in IBFM-2 results, has a dominant component of $f_{7/2}$,
the beta decay rate is small, hence  the $\log_{10}ft$ value of the
$(\frac{9}{2}^{-})_{3}$ in $^{153}$Dy is larger than for the one in
$^{151}$Dy. This enforces our spin assignment in the structure
calculation.

 The $\beta$-decays of ground state $J^{-} =
\frac{11}{2}^{-}$ of $^{151,153}$Ho to $^{151,153}$Dy have many
branches as shown in Table \ref{t12}, but the strong ones are  to
the  level $\frac{9}{2}^{-}_{1}$, which are  60 and 70 percent
\cite{Ric1}. The $\log_{10}ft$ are equal to  ( 4.6 and 3.511) and
(4.7 and 3.612) in the experiments and the model results for  two
isotopes, respectively. In $^{153}$Dy,  with the predicted
assignments $J^{-}(1.092)$ and $J^{-}(1.189)$ = $\frac{11}{2}^{-}$,
the calculated $\log_{10}ft$ of  the two branches are 6.707 and
7.358 respectively, which are close to the experimental value of 6.3
for both branches, respectively.

From Table \ref{t12}, it can be concluded that IBFM-2 provides a
meaningful framework for describing $\beta$-decay transitions
between  $^{151,153}$Dy and $^{151,153}$Ho. Together with other
recent $\beta$-decay calculations
\cite{YoshPRC66,ZufPRC68,BraPRC70}, it shows that the IBFM-2 is well
suited for understanding $\beta$-decay properties   of rare-earth
nuclei.

\section{Summary}
\label{s8}

In this work IBFM-2 calculations for the odd-mass Ho and Dy
(A=151,153) have been presented. For Dy-isotopes four fermion
single-particle  $ f_{7/2}$, $h_{9/2}$, $p_{3/2}$ and $f_{5/2}$
orbitals  were used to study the negative- parity states. For study
the negative- parity states in Ho-isotopes only $h_{11/2}$ has been
taken into account.   The boson-boson interaction parameters were
fixed by the calculation on the boson core nuclei. The boson-fermion
interaction  parameter is kept constant of each element, and  there
are only two free varying boson-fermion quadrupole interaction
parameters for each even-odd nucleus. The analysis  of the wave
functions indicates that the   $ f_{7/2}$ and $h_{9/2}$ orbitals are
dominate in the wave functions  in Dy-isotopes.  The known
quadrupole and magnetic moments in these nuclei are reasonably well
described by the model. The present IBFM-2 calculations provide a
satisfactory framework for  describing the $\beta$-decay rates in
the odd-mass nuclei with $A \sim 150$. The predictions of this work
can serve as a good reference to experimentalists.  Further
experimental study will be very helpful to further test the present
IBFM-2 calculation.

$$\mathbf{Acknowledgements}$$ The authors would like express their thanks to Professor N. Yoshida
for his interest in the subject and his many helpful suggestions.
This work is supported by the National Fundamental Research Program
Grant No. 2006CB921106, China National Natural Science Foundation
Grant Nos. 10325521, 60433050 and the SRFDP program of Education
Ministry of China.


 \begin{table}
\caption{The parameters of the   IBM-2 Hamiltonian.
    $\chi _{\pi} $ =  $\chi _{\nu} $= $-$1.323,  have been
   chosen  for $^{150,152}$Dy and $^{152,154}$Er. All the parameters are in MeV unit.}\label{t1}
\begin{tabular}{cccccccccc}
  \\ \hline\hline
  nucleus &$ \kappa_{\pi\nu} $& $\varepsilon_{d}$ &$ \xi_{1}$ & $\xi_{2}$ &$\xi_{3}$  &$C_{\pi}^{L} (L=0,2,4)$ \\
  \\ \hline\hline
  $^{150}$Dy, $^{152}$Er & $-$0.010 & 0.820 & 0.300 & 0.300 & 0.300 & 0.210,0.210,$-$0.130 \\
  $^{152}$Dy, $^{154}$Er & $-$0.010 & 0.650 &$-$0.400& 0.140 &   0.400 & 0.500,0.500,0.000 \\
   \hline\hline
\end{tabular}
\end{table}

\begin{table} \caption{Single-particle energies (MeV) of proton
orbitals in Ho-isotopes and parameters in the boson-fermion
interaction (MeV).}\label{t2}
\begin{tabular}{cccccccccc}
  \\ \hline\hline
  nucleus &$  g_{9/2}$& $ g_{7/2}$& $d_{5/2}$& $h_{11/2}$& $d_{3/2}$&$s_{1/2}$&$\Gamma$& $A$&$\Lambda$ \\
  \\ \hline\hline
  $^{151}$Ho & $-$5.000 & 0.123 & 0.800 & 2.202 & 3.021 & 3.277&1.200&$-$0.500&0.250 \\
  $^{153}$Ho & $-$5.000 & 0.148 & 0.798 & 2.252 & 3.101 & 3.337&1.200&$-$0.240&0.520 \\
   \hline\hline
\end{tabular}
\end{table}

\begin{table} \caption{Single-particle energies (MeV) of neutron
orbitals in Dy-isotopes and parameters in the boson-fermion
interaction (MeV).}\label{t3}
\begin{tabular}{ccccccccccc}
  \\ \hline\hline
  nucleus &$  h_{11/2}$& $ h_{9/2}$& $f_{7/2}$& $f_{5/2}$& $i_{13/2}$&$p_{3/2}$&$p_{1/2}$&$\Gamma$& $A$&$\Lambda$ \\
  \\ \hline\hline
  $^{151}$Dy & $-$5.800 &  $-$0.4300 & $-$1.450 & 1.760 &1.350& 1.450&2.740&0.500 &$-$0.300&0.350 \\
  $^{153}$Dy & $-$5.800 & $-$0.450 & $-$1.500 & 1.800 & 1.350&1.500 & 2.800&0.500&$-$0.440&0.550 \\
   \hline\hline
\end{tabular}
\end{table}


\begin{table}
\begin{center}
\caption{ Available experimental and calculated energy levels for
$^{153}$Dy isotope. For details calculation  see figure
\ref{f8}.}\label{t4}
\begin{tabular}{ccccccccc}\hline\hline
 $\underline{IBFM-2}$ & &    $\underline{EXP.}$    \\
   $J^{-}$& Energy   &$ J^{-}$
  &Energy\\\hline\hline
$7^{-}/2$&0.000&$7^{-}/2$&0.000\\
$3^{-}/2$&0.104&$3^{-}/2$&0.108\\
$5^{-}/2$&0.327&$3^{-}/2,5^{-}/2$&0.270\\
$9^{-}/2$&0.331&$9^{-}/2$&0.295\\
$1^{-}/2$&0.438&(-)&0.500\\
$9^{-}/2$&0.600&$5^{-}/2,7^{-}/2,9^{-}/2$&0.365\\
 $7^{-}/2$&0.606&$3^{-}/2,5^{-}/2,7^{-}/2$&0.576\\
 $5^{-}/2$&0.646&$3^{-}/2,5^{-}/2$&0.565\\
 $11^{-}/2$&0.657&$11^{-}/2$&0.637\\
 $7^{-}/2$&0.695&$5^{-}/2,7^{-}/2$&0.688\\
 $7^{-}/2$&0.879&$7^{-}/2$&0.830\\
 $13^{-}/2$&0.877&$13^{-}/2$&0.837\\
  $13^{-}/2$&1.195&$13^{-}/2$&1.041\\
  $11^{-}/2$&1.151&$11^{-}/2$&1.068\\
  $11^{-}/2$&1.236&$9^{-}/2,11^{-}/2$&1.092\\
  $11^{-}/2$&1.252&$7^{-}/2,9/2,11^{-}/2$&1.189\\
  $15^{-}/2$&1.254&$15^{-}/2$&1.273\\
  $13^{-}/2$&1.277&$13^{-}/2$&1.321\\
  $17^{-}/2$&1.394&$17^{-}/2$&1.455\\
  $15^{-}/2$&1.543&$15^{-}/2$&1.584\\
  $17^{-}/2$&1.752&$17^{-}/2$&1.602\\
  $19^{-}/2$&1.796&$19^{-}/2,17^{-}/2$&1.753\\
  $17^{-}/2$&1.846&$17^{-}/2$&1.862\\
  $21^{-}/2$&1.892&$21^{-}/2$&2.043\\
  $19^{-}/2$&1.922&$19^{-}/2$&2.150\\
  $21^{-}/2$&2.288&$21^{-}/2$&2.454\\
  $25^{-}/2$&2.621&$25^{-}/2$&2.523\\\hline

 \end{tabular}
\end{center}
\end{table}

\begin{table}
\begin{center}
\caption{ The percentage components  of wave functions for
$^{151,153}$Dy-isotopes.} \label{t5}
\begin{tabular}{ccccccccc}\hline\hline
  &  \underline{}  & \underline{$^{151}$Dy}&& &&\underline{$^{153}$Dy}\\
    $J^{-}$              &$p_{3/2}$&$f_{7/2}$& $h_{9/2}$& $f_{5/2}$&
    \hspace{.052in}$p_{3/2}$&$f_{7/2}$& $h_{9/2}$& $f_{5/2}$\\ \hline\hline
 $7/2^{ -}_ { 1}  $ &    2& 97&  0&  0&\hspace{.052in}  2 &97 &1 &0  \\
 $3/2^{ -}_ { 1}  $ &    13& 86&  1&  1&\hspace{.052in}  9 &90&0&1   \\
 $5/2^{ -}_ { 1}  $ &    2& 95& 1&  1&\hspace{.052in} 2&80&17&1      \\
 $9/2^{ -}_ { 1}  $ &    0&  6& 90&  4&\hspace{.052in} 0&1&94&4      \\
 $1/2^{ -}_ { 1}  $ &   18& 78&  2&  1&\hspace{.052in} 15&82&2&1     \\
 $9/2^{ -}_ { 2}  $ &    1& 91&  7&  0&\hspace{.052in} 1&93&5&0      \\
 $7/2^{ -}_ { 2}  $ &   4& 95&  1&  0& \hspace{.052in} 8&90&1&1      \\
 $5/2^{ -}_ { 2}  $ &    1& 4& 81&  14&\hspace{.052in}4&53&38&5      \\
 $11/2^{ -}_ { 1} $  &    5& 93&  1&  0&\hspace{.052in} 3&95&1&0     \\
 $7/2^{ -}_ { 3}  $ &    0& 5& 90&  4&\hspace{.052in} 3&46&49&2      \\
 $5/2^{ -}_ { 3}  $&    13& 83& 4&  1&\hspace{.052in} 8&64&22&6      \\
 $3/2^{ -}_ { 2}  $&    12& 85&  1&  1& \hspace{.052in}8&88&3&1      \\
 $9/2^{ -}_ { 3}  $&    1& 14&  81&  4& \hspace{.052in} 4&92&4&0      \\
$ 3/2^{ -}_ { 3} $ &    1& 10& 75&  14& \hspace{.052in}  13&77&9&2    \\
$11/2^{ -}_ { 2} $&    0&  8& 89&  3& \hspace{.052in} 5&91&3&1       \\
$13/2^{ -}_ { 1} $&    0&  6& 87&  7&\hspace{.052in} 0&1&93&6        \\
$ 1/2^{ -}_ { 2} $ &    1& 5& 78&  17& \hspace{.052in}10&50&34&7       \\
$11/2^{ -}_ { 3} $&    5& 88&  6&  1& \hspace{.052in} 0&3&93&3       \\
$ 13/2 ^{ -}_ { 2}   $ &3&  87& 9& 1& \hspace{.052in} 0&6&89&5       \\
$ 15/2 ^{ -}_ { 1}   $ &10& 88&  1& 1&  \hspace{.052in}5&94&1&0       \\
$ 17/2 ^{ -}_ { 1}   $ &1&  8& 82& 9&  \hspace{.052in}0&0&92&8       \\
$ 19/2 ^{ -}_ { 1}   $ &16& 81&  2& 1& \hspace{.052in} 60&39&0&1       \\
$ 21/2^{ -}_ { 1}  $   &3& 13&  74&10&  \hspace{.052in}0&0&91&9       \\
$ 23/2^{ -}_ {  1}  $  &24&73 &  2& 1& \hspace{.052in}57&41&1&1        \\
$ 25/2 ^{ -}_ {  1}   $&15&  32& 45&8& \hspace{.052in}  1&1&88&10     \\
\\ \hline
\end{tabular}
\end{center}
\end{table}

\begin{table}
\begin{center}
\caption{ Branching ratios in $^{151}$Ho isotope, the order of
states corresponding to IBFM-2 calculation.}\label{t6}
\begin{tabular}{cccccccc}\hline\hline
Level (MeV)& Transition & $I_{\gamma}$(IBFM-2)&$I_{\gamma}$(Exp.)
 \\\hline\hline
0.638& $9^{-}/2_{1}\rightarrow 11^{-}/2_{1}$&100 &100\\

0.667& $7^{-}/2_{1}\rightarrow 9^{-}/2_{1}$&0.1 &\\
     & $7^{-}/2_{1}\rightarrow 11^{-}/2_{1}$&100 &100\\

0.789& $15^{-}/2_{1}\rightarrow 11^{-}/2_{1}$&100 &100\\

0.869& $13^{-}/2_{1}\rightarrow 15^{-}/2_{1}$&0.2 &\\
     & $13^{-}/2_{1}\rightarrow 9^{-}/2_{1}$&0.0 & 9.7(12)\\
     & $13^{-}/2_{1}\rightarrow 11^{-}/2_{1}$&100 &100(9)\\

0.910& $11^{-}/2_{2}\rightarrow 13^{-}/2_{1}$&0.0 &\\
     & $11^{-}/2_{2}\rightarrow 15^{-}/2_{1}$&0.0 &  \\
     & $11^{-}/2_{2}\rightarrow 7^{-}/2_{1}$&0.0 & \\
     & $11^{-}/2_{2}\rightarrow 9^{-}/2_{1}$&7.3 & \\
     & $11^{-}/2_{2}\rightarrow 11^{-}/2_{1}$&100 &\\

1.129&$5^{-}/2_{1}\rightarrow 7^{-}/2_{1}$&100& 38(3) \\
     & $7^{-}/2_{2}\rightarrow 9^{-}/2_{1}$&43.7 & \\

1.279& $7^{-}/2_{2}\rightarrow 5^{-}/2_{1}$&25.4 &\\
     & $7^{-}/2_{2}\rightarrow 11^{-}/2_{1}$&0.6 & \\
     & $7^{-}/2_{2}\rightarrow 7^{-}/2_{1}$&69.4 &\\
     &$7^{-}/2_{2}\rightarrow 9^{-}/2_{1}$&100 &100\\
     &$7^{-}/2_{2}\rightarrow 11^{-}/2_{1}$&10.3 &\\
1.387& $19^{-}/2_{1}\rightarrow 15^{-}/2_{1}$&100 &100\\

1.541& $7^{-}/2_{3}\rightarrow 5^{-}/2_{1}$&0.0 &\\
     & $7^{-}/2_{3}\rightarrow 7^{-}/2_{2}$&0.0&  \\
     & $7^{-}/2_{3}\rightarrow 11^{-}/2_{2}$&0.5& \\
     & $7^{-}/2_{3}\rightarrow 7^{-}/2_{1}$&100 &100 \\
     & $7^{-}/2_{3}\rightarrow 9^{-}/2_{1}$&85.5 & \\
     & $7^{-}/2_{3}\rightarrow 11^{-}/2_{1}$&70.6 & \\
\\\hline
\end{tabular}
\end{center}
\end{table}

\begin{table}
\begin{center}
\caption{ Branching ratios in $^{153}$Ho isotope, the order of
states corresponding to IBFM-2 calculation.}\label{t7}
\begin{tabular}{cccccccc}\hline\hline
Level (MeV)& Transition & $I_{\gamma}$(IBFM-2)&$I_{\gamma}$(Exp.)
 \\\hline\hline
0.351& $9^{-}/2_{1}\rightarrow 11^{-}/2_{1}$&100 &100\\

0.398& $7^{-}/2_{1}\rightarrow 9^{-}/2_{1}$&7.9 & \\
     & $7^{-}/2_{1}\rightarrow 11^{-}/2_{1}$&100 &100\\

0.576& $15^{-}/2_{1}\rightarrow 11^{-}/2_{1}$&100 &100\\

0.706& $13^{-}/2_{1}\rightarrow 15^{-}/2_{1}$&0.7 & \\
     & $13^{-}/2_{1}\rightarrow 9^{-}/2_{1}$&0.3 &\\
     & $13^{-}/2_{1}\rightarrow 11^{-}/2_{1}$&100 &\\

0.814& $7^{-}/2_{2}\rightarrow 7^{-}/2_{1}$&0.5 & \\
     & $7^{-}/2_{2}\rightarrow 9^{-}/2_{1}$&100 &100\\
     & $7^{-}/2_{2}\rightarrow 11^{-}/2_{1}$&0.1 &\\

 0.926& $9^{-}/2_{2}\rightarrow 7^{-}/2_{2}$&4.8 & \\
     & $9^{-}/2_{2}\rightarrow13^{-}/2_{1}$&0.0&\\
     & $9^{-}/2_{2}\rightarrow 7^{-}/2_{1}$&62.5 & \\
     & $9^{-}/2_{2}\rightarrow 9^{-}/2_{1}$&54.7 &100\\
1.207& $19^{-}/2_{1}\rightarrow 15^{-}/2_{1}$&100&100 \\

\\\hline

 \end{tabular}
\end{center}
\end{table}

\begin{table}
\begin{center}
\caption{ Branching ratios in $^{151}$Dy isotope, the order of
states corresponding to IBFM-2 calculation.}\label{t8}
\begin{tabular}{cccccccc}\hline\hline
Level (MeV)& Transition & $I_{\gamma}$(IBFM-2)&$I_{\gamma}$(Exp.)
 \\\hline\hline
0.527& $9^{-}/2_{1}\rightarrow 7^{-}/2_{1}$&100 &100\\

0.775& $11^{-}/2_{1}\rightarrow 9^{-}/2_{1}$&5.1 &\\
     & $11^{-}/2_{1}\rightarrow 7^{-}/2_{1}$&100 &100\\

0.984& $9^{-}/2_{2}\rightarrow 11^{-}/2_{1}$&5.0 &100\\
     & $9^{-}/2_{2}\rightarrow 9^{-}/2_{1}$&0.1 &\\
     & $9^{-}/2_{2}\rightarrow 7^{-}/2_{1}$&100 & \\
1.334& $11^{-}/2_{2}\rightarrow 9^{-}/2_{2}$&0.1 &22(13)\\
      & $11^{-}/2_{2}\rightarrow 11^{-}/2_{1}$&0.0 &50(13)\\
     & $11^{-}/2_{2}\rightarrow 9^{-}/2_{1}$&100 &  \\
     & $11^{-}/2_{2}\rightarrow 7^{-}/2_{1}$&0.2 &    \\

1.348& $13^{-}/2_{1}\rightarrow 11^{-}/2_{2}$&0.0 &\\
     &  $13^{-}/2_{1}\rightarrow 9^{-}/2_{2}$&0.0 &\\
     & $13^{-}/2_{1}\rightarrow 11^{-}/2_{1}$&0.8 &7.5(15)\\
     & $13^{-}/2_{1}\rightarrow 9^{-}/2_{1}$&100 &100(2) \\

1.511& $15^{-}/2_{1}\rightarrow 13^{-}/2_{1}$&1.7 &\\
       & $15^{-}/2_{1}\rightarrow 11^{-}/2_{2}$&0.0 &\\
     & $15^{-}/2_{1}\rightarrow 11^{-}/2_{1}$&100 &100(3)\\

1.549& $9^{-}/2_{3}\rightarrow 13^{-}/2_{1}$&0.0 &\\
      & $9^{-}/2_{3}\rightarrow 11^{-}/2_{2}$&5.6 &\\
     & $9^{-}/2_{3}\rightarrow 9^{-}/2_{2}$&0.3&\\
     & $9^{-}/2_{3}\rightarrow 11^{-}/2_{1}$&23.9 & \\
     & $9^{-}/2_{3}\rightarrow 9^{-}/2_{1}$&100 &36(8)\\
     & $9^{-}/2_{3}\rightarrow 7^{-}/2_{1}$&3.5 & 100(7)\\

1.918& $17^{-}/2_{1}\rightarrow 15^{-}/2_{1}$&0.6 &43.2(13)\\
     & $17^{-}/2_{1}\rightarrow 13^{-}/2_{1}$&100 &100(2)\\

2.263& $21^{-}/2_{1}\rightarrow 17^{-}/2_{1}$&100&100\\

\\\hline
 \end{tabular}
\end{center}
\end{table}

\begin{table}
\begin{center}
\caption{ Branching ratios in $^{153}$Dy isotope, the order of
states corresponding to IBFM-2 calculation.}\label{t9}
\begin{tabular}{ccccccccc}\hline\hline
Level (MeV)& Transition & $I_{\gamma}$(IBFM-2)&$I_{\gamma}$(Exp.)
 \\\hline\hline
0.108& $3^{-}/2_{1}\rightarrow 7^{-}/2_{1}$&100 &100\\
0.270& $5^{-}/2_{1}\rightarrow 3^{-}/2_{1}$&100 &100(5)\\
     & $5^{-}/2_{1}\rightarrow 7^{-}/2_{1}$&20.2 &86(4)\\

0.295& $9^{-}/2_{1}\rightarrow 5^{-}/2_{1}$&0.0 &\\
     & $9^{-}/2_{1}\rightarrow 7^{-}/2_{1}$&100 &100\\

0.365& $9^{-}/2_{2}\rightarrow 9^{-}/2_{1}$&0.2 &\\
     & $9^{-}/2_{2}\rightarrow 5^{-}/2_{1}$&0.0 &$\approx3.3$\\
     & $9^{-}/2_{2}\rightarrow 7^{-}/2_{1}$&100 &100(27)\\

0.565& $5^{-}/2_{2}\rightarrow 9^{-}/2_{2}$&0.0&\\
     & $5^{-}/2_{2}\rightarrow 9^{-}/2_{1}$&2.0 &\\
     & $5^{-}/2_{2}\rightarrow 5^{-}/2_{1}$&11.3 &  \\
     & $5^{-}/2_{2}\rightarrow 3^{-}/2_{1}$&50.9 &100(4)\\
     & $5^{-}/2_{2}\rightarrow 7^{-}/2_{1}$&100 &48(15)\\

0.576& $7^{-}/2_{2}\rightarrow 5^{-}/2_{2}$&0.0 &\\
     & $7^{-}/2_{2}\rightarrow 9^{-}/2_{2}$&38.5 &\\
     & $7^{-}/2_{2}\rightarrow 9^{-}/2_{1}$&1.4 &\\
     & $7^{-}/2_{2}\rightarrow 5^{-}/2_{1}$&100 &\\
     & $7^{-}/2_{2}\rightarrow 3^{-}/2_{1}$&30.7 &88(15)\\
     & $7^{-}/2_{2}\rightarrow 7^{-}/2_{1}$&72.9 &100(19)\\

0.637& $11^{-}/2_{1}\rightarrow 7^{-}/2_{2}$&0.0 &\\
     & $11^{-}/2_{1}\rightarrow 9^{-}/2_{2}$&71.6 &\\
     & $11^{-}/2_{1}\rightarrow 9^{-}/2_{1}$&19.3 &1.2\\
     & $11^{-}/2_{1}\rightarrow 7^{-}/2_{1}$&100 &100(4)\\

0.688& $7^{-}/2_{4}\rightarrow 7^{-}/2_{2}$&7.1 &\\
     & $7^{-}/2_{4}\rightarrow 5^{-}/2_{2}$&38.8 &\\
     & $7^{-}/2_{4}\rightarrow 9^{-}/2_{2}$&75.1 &\\
     & $7^{-}/2_{4}\rightarrow 9^{-}/2_{1}$&100.0 &9(4)\\
     & $7^{-}/2_{4}\rightarrow 5^{-}/2_{1}$&41.7 &\\
     & $7^{-}/2_{4}\rightarrow 3^{-}/2_{1}$&40.3 &22(9)\\
     & $7^{-}/2_{4}\rightarrow 7^{-}/2_{1}$&38.3 &100(7)\\

0.837& $13^{-}/2_{1}\rightarrow 11^{-}/2_{1}$&0.1 &\\
     & $13^{-}/2_{1}\rightarrow 9^{-}/2_{2}$&0.1 &\\
     & $13^{-}/2_{1}\rightarrow 9^{-}/2_{1}$&100 &100(14)\\

1.041& $13^{-}/2_{2}\rightarrow 13^{-}/2_{1}$&63.2 &$\approx$50\\
     & $13^{-}/2_{2}\rightarrow 11^{-}/2_{1}$&1.3 &100(50)\\
     & $13^{-}/2_{2}\rightarrow 9^{-}/2_{2}$&0.0 &\\
     & $13^{-}/2_{2}\rightarrow 9^{-}/2_{1}$&100 &100(50)\\

1.273& $15^{-}/2_{1}\rightarrow 13^{-}/2_{2}$& 17.2 & \\
     & $15^{-}/2_{1}\rightarrow 13^{-}/2_{1}$& 0.3& \\
     & $15^{-}/2_{1}\rightarrow 11^{-}/2_{1}$&100 &100(24)\\

\\\hline

 \end{tabular}
\end{center}
\end{table}

\begin{table}
\begin{center}
\caption{Experimental and calculated B(E2) (in unit $e^{2}b^{2}$)
and  B(M1) (in unit $\mu_{N}^{2}) $, the Quadrupole moment and
Magnetic moment  of ground state and low-lying  states listed in
last lines  for $^{151,153}$Ho isotopes.}\label{t10}
\begin{tabular}{ccccccccccc}\hline\hline
 &  & \underline{  $^{151}$Ho  }  &    &  &    & \underline{  $^{153}$Ho      }   &      &  &  \\
      &\hspace{ 0.3in}\underline{  B(E2) } &   &   \hspace{ 0.3in} \underline{ B(M1) }&  &
       \hspace{ 0.3in}  \underline{ B(E2) } &   & \hspace{ 0.3in} \underline{ B(M1) }&  &  \\
  $J_{i}^{-}\rightarrow  J_{f}^{-}$& IBFM-2 & EXP. & IBFM-2 & EXP.&IBFM-2 & EXP. & IBFM-2 & EXP. \\  \hline\hline
  $9^{-}/2_{1}\rightarrow11^{-}/2_{1}$ & 0.0612 &       & 0.0216       &        & 0.0628&        &  0.0684    &   \\

  $7^{-}/2_{1}\rightarrow11^{-}/2_{1}$ & 0.0645 &       &         &       &0.0596&        &        &    \\
  $7^{-}/2_{1}\rightarrow9^{-}/2_{1}$ & 0.0150 &       & 0.2398  &        & 0.0204&       &0.2956      &     \\
  $15^{-}/2_{1}\rightarrow11^{-}/2_{1}$ & 0.0659 &       &        &        & 0.0643&       &       &     \\
  $13^{-}/2_{1}\rightarrow11^{-}/2_{1}$ & 0.0697 &       &0.0083  &        & 0.0717&       & 0.0013&     \\

 $13^{-}/2_{1}\rightarrow9^{-}/2_{1}$ & 0.0082&       &          &        &0.0076&        &       &    \\
 $13^{-}/2_{1}\rightarrow15^{-}/2_{1}$ & 0.0027 &        & 0.0887 &        & 0.0061&       &0.0274 &     \\
 $11^{-}/2_{2}\rightarrow11^{-}/2_{1}$ & 0.0664 &       &0.0005 &        & 0.0650&         & 0.0006&     \\
 $11^{-}/2_{2}\rightarrow9^{-}/2_{1}$ & 0.0039 &       &0.1046  &        & 0.0006&       & 0.0103&     \\

 $11^{-}/2_{2}\rightarrow13^{-}/2_{1}$ & 0.0004 &       & 0.0942  &        &0.0015&        & 0.0153&    \\
$11^{-}/2_{2}\rightarrow15^{-}/2_{1}$ & 0.0039 &        &         &        & 0.0036&       &       &     \\
$5^{-}/2_{1}\rightarrow7^{-}/2_{1}$ & 0.0865 &         &0.0041  &        & 0.0850&       & 0.0133&     \\
$5^{-}/2_{1}\rightarrow9^{-}/2_{1}$ & 0.0363 &       &       &        & 0.0345&         &        &     \\

$7^{-}/2_{2}\rightarrow7^{-}/2_{1}$ & 0.0449 &       & 0.0007&        &0.0290&        & 0.0002&    \\
$7^{-}/2_{2}\rightarrow5^{-}/2_{1}$ & 0.0080 &        & 0.4656  &        & 0.0093&       &0.7639 &     \\
$3^{-}/2_{1}\rightarrow7^{-}/2_{1}$ & 0.1238&       &      &        & 0.1093&         &      &     \\
$3^{-}/2_{1}\rightarrow7^{-}/2_{2}$ & 0.0249  &       &     &        & 0.0182&         &       &     \\

$13^{-}/2_{2}\rightarrow11^{-}/2_{2}$ & 0.0071 &       & 0.0019&        & 0.0005&         & 0.0022 &     \\
$13^{-}/2_{2}\rightarrow13^{-}/2_{1}$ & 0.0131 &       & 0.0018&        & 0.0046&         &  0.0306 &     \\
$9^{-}/2_{2}\rightarrow7^{-}/2_{2}$ & 0.0015 &          &0.4275 &        & 0.0010&         &0.1388        &     \\
$9^{-}/2_{2}\rightarrow9^{-}/2_{1}$ & 0.0625 &       & 0.0035&        & 0.0713&         & 0.0077  &     \\

$11^{-}/2_{1}$                        & $-$0.2858 &        & 6.9255&        & $-$0.2303& $-$1.1(5)& 6.8612& 6.81 \\
$9^{-}/2_{1}$                        &  $-$0.0546 &        & 5.9162&         & $-$0.0359&         & 5.9166&     \\
$7^{-}/2_{1}$                        & $-$0.2761 &        & 5.4478&         & $-$0.2538&         & 5.7557&      \\
$15^{-}/2_{1}$                        & $-$0.4559 &        & 8.2781&         & $-$0.4735&       & 7.9279&      \\
$13^{-}/2_{1}$                        & $-$0.2215 &        & 7.6754&         &$-$0.1816&       & 7.7220&        \\
$11^{-}/2_{2}$                        & $-$0.0582 &        & 6.7618&         &$-$0.0271&       & 6.7609&      \\
$7^{-}/2_{2}$                        &   0.1773 &        & 4.5124&         & 0.2187&       & 4.6364&       \\
$5^{-}/2_{1}$                        &  0.1045&        & 3.7040&         & 0.1267&        & 4.0168&      \\
$3^{-}/2_{1}$                        & $-$0.1572 &        & 3.1244&         & $-$0.1507&        & 3.8207&       \\
\hline\hline\\
\end{tabular}
\end{center}
\end{table}

\begin{table}
\begin{center}
\caption{Experimental and calculated B(E2) (in unit $e^{2}b^{2}$)
and  B(M1) (in unit $\mu_{N}^{2}) $, the Quadrupole moment and
Magnetic moment  of ground state and low-lying  states listed in
last lines  for $^{151,153}$Dy isotopes.}\label{t11}
\begin{tabular}{ccccccccccc}\hline\hline
 &  & \underline{  $^{151}$Dy  }  &    &  &    & \underline{  $^{153}$Dy      }   &      &  &  \\
      &\hspace{ 0.3in}\underline{  B(E2) } &   &   \hspace{ 0.3in} \underline{ B(M1) }&  &
       \hspace{ 0.3in}  \underline{ B(E2) } &   & \hspace{ 0.3in} \underline{ B(M1) }&  &  \\
  $J_{i}^{-}\rightarrow  J_{f}^{-}$& IBFM-2 & EXP. & IBFM-2 & EXP.&IBFM-2 & EXP. & IBFM-2 & EXP. \\  \hline\hline
  $3^{-}/2_{1}\rightarrow7^{-}/2_{1}$ & 0.1173 &       &        &        & 0.0389&0.9135(728)&      &   \\

  $9^{-}/2_{1}\rightarrow7^{-}/2_{1}$ & 0.0079 &       & 0.0008  &       &0.0019&        & 0.0002&    \\
  $11^{-}/2_{1}\rightarrow7^{-}/2_{1}$ & 0.1022 &       &        &        & 0.0364&       &      &     \\
  $5^{-}/2_{1}\rightarrow7^{-}/2_{1}$ & 0.0738 &       & 0.1245  &        & 0.0470&$>$0.0534& 0.0016&     \\
  $11^{-}/2_{1}\rightarrow9^{-}/2_{1}$ & 0.0009 &       &0.0668  &        & 0.0008&       & 0.0128&     \\

 $5^{-}/2_{1}\rightarrow3^{-}/2_{1}$ & 0.0140 &       & 0.6252  &        &0.0001&        & 0.0910&    \\
 $5^{-}/2_{1}\rightarrow9^{-}/2_{1}$ & 0.0050 &        &        &        & 0.0221&       &      &     \\
 $9^{-}/2_{2}\rightarrow7^{-}/2_{1}$ & 0.0640 &       & 0.0417  &        & 0.0533&         & 0.0019&     \\
 $7^{-}/2_{2}\rightarrow7^{-}/2_{1}$ & 0.0671 &       &0.0166  &        & 0.0147&       & 0.0051&     \\

 $7^{-}/2_{2}\rightarrow5^{-}/2_{1}$ & 0.0094 &       & 0.7565  &        &0.0036&        & 0.0778&    \\
$7^{-}/2_{2}\rightarrow9^{-}/2_{1}$ & 0.0001 &        & 0.0384  &        & 0.0002&       &0.0014 &     \\
$7^{-}/2_{2}\rightarrow9^{-}/2_{2}$ & 0.0002 &       &0.7781  &        & 0.0012&       & 0.0915&     \\
$9^{-}/2_{2}\rightarrow9^{-}/2_{1}$ & 0.0028 &       & 0.0005  &        & 0.0022&         & 0.0016&     \\

$5^{-}/2_{2}\rightarrow7^{-}/2_{1}$ & 0.0004 &       & 0.0335 &        &0.0149&        & 0.0153&    \\
$5^{-}/2_{2}\rightarrow7^{-}/2_{2}$ & 0.0010 &        & 0.0040  &        & 0.0059&       &0.0065 &     \\
$5^{-}/2_{2}\rightarrow3^{-}/2_{1}$ & 0.0048 &       &0.0001  &        & 0.0063&       & 0.0170&     \\
$11^{-}/2_{2}\rightarrow7^{-}/2_{2}$ & 0.0028 &       &      &        & 0.0015&         &      &     \\
$11^{-}/2_{2}\rightarrow9^{-}/2_{1}$ & 0.0731 &       & 0.0554&        & 0.0014&         & 0.0043 &     \\
$11^{-}/2_{2}\rightarrow9^{-}/2_{2}$ & 0.0016 &       & 0.0006&        & 0.0018&         & 0.1213 &     \\
$13^{-}/2_{1}\rightarrow9^{-}/2_{1}$ & 0.1174 &       &       &        & 0.1030&         &        &     \\
$13^{-}/2_{1}\rightarrow11^{-}/2_{1}$ & 0.0047 &       & 0.0003&        & 0.0000&         &  0.0006 &     \\
$15^{-}/2_{1}\rightarrow13^{-}/2_{1}$ & 0.0012 &       & 0.1076&        & 0.0001&         & 0.0001  &     \\
$7^{-}/2_{1}$                      & $-$0.3868 & $-$0.30(5)& $-$0.8439& $-$0.945(7)& $-$0.1410& $-$0.02(5)& $-$0.8907& -0.782(6) \\
$9^{-}/2_{1}$                        & $-$0.6378 &      & 0.9640&         & $-$0.6726&      & 0.9553&      \\
$3^{-}/2_{1}$                        & $-$0.2672 &     &  $-$1.7209&         & $-$0.2167&      & $-$1.2199&      \\
$11^{-}/2_{1}$                       & $-$0.6098 &     & 1.0877  &         & $-$0.3413&      &  0.4119&      \\
$5^{-}/2_{1}$                        & $-$0.0568 &     & $-$0.6930&         & $-$0.0712&       & $-$0.6369&      \\
$9^{-}/2_{2}$                        & $-$0.3249 &     &  0.4675&         & $-$0.1749&        & $-$0.1302&      \\
$7^{-}/2_{2}$                        & $-$0.0702 &     & $-$0.0038&         & $-$0.3029&        & 0.1622&      \\
$5^{-}/2_{2}$                        & $-$0.5281 &     & $-$0.2779&         & $-$0.1388&        & $-$0.2006&      \\
\hline\hline\\
\end{tabular}
\end{center}
\end{table}

\begin{table}
\begin{center}
\caption{ Comparison of $log_{10}ft$ values in the decays $^{151}$Ho
$\rightarrow$ $^{151}$Dy and$^{153}$Ho $\rightarrow$ $^{153}$Dy, the
order of states corresponding to IBFM-2 calculation.}\label{t12}
\begin{tabular}{ccccccccccccccccccccc}\hline\hline\\
$^{151}$Ho &$\rightarrow$& $^{151}$Dy  &&&&$^{153}$Ho
&$\rightarrow$& $^{153}$Dy
\\
odd-p&  & odd-n& log ft &  Exp. &&  odd-p&  & odd-n& log ft &
Exp.                            \\\hline\hline
 $ 11^{-}/2_{1}$ &$\rightarrow $&$13^{-}/2_{1}$& 7.260&6.200&& $ 11^{-}/2_{1}$ &$\rightarrow $&$13^{-}/2_{1}$& 7.231&6.300\\
$ 11^{-}/2_{1}$ &$\rightarrow $&$13^{-}/2_{2}$& 8.440 &  &&$ 11^{-}/2_{1}$ &$\rightarrow $&$13^{-}/2_{2}$& 7.793&     \\
$ 11^{-}/2_{1}$ &$\rightarrow $&$13^{-}/2_{3}$& 8.907&   &&$ 11^{-}/2_{1}$ &$\rightarrow $&$13^{-}/2_{3}$& 9.456& 6.200     \\
$ 11^{-}/2_{1}$ &$\rightarrow $&$13^{-}/2_{4}$& 9.174&&&$ 11^{-}/2_{1}$ &$\rightarrow $&$13^{-}/2_{4}$& 10.228&  \\
$ 11^{-}/2_{1}$ &$\rightarrow $&$13^{-}/2_{5}$& 8.807&   &&$ 11^{-}/2_{1}$ &$\rightarrow $&$13^{-}/2_{5}$& 8.515&      \\
$ 11^{-}/2_{1}$ &$\rightarrow $&$11^{-}/2_{1}$& 8.019&&&$ 11^{-}/2_{1}$ &$\rightarrow $&$11^{-}/2_{1}$& 7.961&6.000\\
$ 11^{-}/2_{1}$ &$\rightarrow $&$11^{-}/2_{2}$& 5.429&5.500   &&$ 11^{-}/2_{1}$ &$\rightarrow $&$11^{-}/2_{2}$& 7.210&     \\
$ 11^{-}/2_{1}$ &$\rightarrow $&$11^{-}/2_{3}$& 6.877&   &&$ 11^{-}/2_{1}$ &$\rightarrow $&$11^{-}/2_{3}$& 5.366&     \\
$ 11^{-}/2_{1}$ &$\rightarrow $&$11^{-}/2_{4}$& 7.742&&&$ 11^{-}/2_{1}$ &$\rightarrow $&$11^{-}/2_{4}$& 9.596&\\
$ 11^{-}/2_{1}$ &$\rightarrow $&$11^{-}/2_{5}$& 7.195&&&$ 11^{-}/2_{1}$ &$\rightarrow $&$11^{-}/2_{5}$&6.707&6.300 \\
$ 11^{-}/2_{1}$ &$\rightarrow $&$11^{-}/2_{6}$& 6.753& &&$ 11^{-}/2_{1}$ &$\rightarrow $&$11^{-}/2_{6}$& 7.358&6.300\\
$ 11^{-}/2_{1}$ &$\rightarrow $&$9^{-}/2_{1}$& 3.511&4.600&&$ 11^{-}/2_{1}$ &$\rightarrow $&$9^{-}/2_{1}$& 3.612&4.700\\
$ 11^{-}/2_{1}$ &$\rightarrow $&$9^{-}/2_{2}$& 4.759&5.800&&$ 11^{-}/2_{1}$ &$\rightarrow $&$9^{-}/2_{2}$& 7.175&6.500\\
$ 11^{-}/2_{1}$ &$\rightarrow $&$9^{-}/2_{3}$& 5.208& 5.100   &&$ 11^{-}/2_{1}$ &$\rightarrow $&$9^{-}/2_{3}$& 6.443&    \\
$ 11^{-}/2_{1}$ &$\rightarrow $&$9^{-}/2_{4}$& 7.320&    &&$ 11^{-}/2_{1}$ &$\rightarrow $&$9^{-}/2_{4}$& 4.889&    \\
$ 11^{-}/2_{1}$ &$\rightarrow $&$9^{-}/2_{5}$&  5.506&  &&$11^{-}/2_{1}$ &$\rightarrow $&$9^{-}/2_{5}$&6.570&
\\\hline
\hline\\

\end{tabular}
\end{center}
\end{table}

\end{document}